# Radio Resource Allocation for Scalable Video Services over Wireless Cellular Networks

Mostafa Zaman Chowdhury[a], Tuan Nguyen[a], Young-Il Kim[b], Won Ryu[b], and Yeong Min Jang[a]

[a]Department of Electronics Engineering, Kookmin University, Seoul, Korea.
[b]Electronics and Telecommunications Research Institute (ETRI), Korea

E-mail: mzceee@yahoo.com, yjang@kookmin.ac.kr

**Abstract** Good quality video services always require higher bandwidth. Hence, to provide the video services e.g., multicast/broadcast services (MBS) and unicast services along with the existing voice, internet, and other background traffic services over the wireless cellular networks, it is required to efficiently manage the wireless resources in order to reduce the overall forced call termination probability, to maximize the overall service quality, and to maximize the revenue. Fixed bandwidth allocation for the MBS sessions either reduces the quality of the MBS videos and bandwidth utilization or increases the overall forced call termination probability and of course the handover call dropping probability as well. Scalable Video Coding (SVC) technique allows the variable bit rate allocation for the video services. In this paper, we propose a bandwidth allocation scheme that efficiently allocates bandwidth among the MBS sessions and the non-MBS traffic calls (e.g., voice, unicast, internet, and other background traffic). The proposed scheme reduces the bandwidth allocation for the MBS sessions during the congested traffic condition only to accommodate more calls in the system.  Instead of allocating fixed bandwidths for the BMS sessions and the non-MBS traffic, our scheme allocates variable bandwidths for them. However, the minimum quality of the videos is guaranteed by allocating minimum bandwidth for them. Using the mathematical and numerical analyses, we show that the proposed scheme maximizes the bandwidth utilization and significantly reduces the overall forced call termination probability as well as the handover call dropping probability.

**Keywords** *MBS, scalable, QoS, bandwidth allocation, video session, call dropping probability, call blocking probability, and bandwidth utilization.*

## 1. Introduction

Video over wireless networks is one of the fastest-growing data applications and mobile TV has become popular as it promises to deliver video contents to users whenever they want and wherever they are. Mobile TV has already proved to be a very promising ARPU (average revenue per user) generator for cellular operators [1]. In case of mobile TV, users can enjoy video services anywhere even with full mobility support through the several access networks. Worldwide Interoperability for Microwave Access (WiMAX) is a typical example of an emerging wireless network system to support high data rate services. The Mobile WiMAX (802.16e) is capable of providing high data rate with quality of service (QoS) mechanisms and making the support of mobile TV very attractive [2]. Therefore, with the rapid improvement of various wireless network technologies, now it is possible to provide high quality video transmission along with the voice, internet, and other background traffic over the wireless networks. Most of the existing wireless network technologies such as femtocell [3], WiFi, Mobile WiMAX, 3G, and 4G support multicast/broadcast mechanisms [4-8]. However, the wireless bandwidth is still insufficient to support huge voice, data, and video services with full QoS especially for the congested traffic condition.

   Good quality video services always require higher bandwidth. Hence, to provide the video services e.g., multicast/broadcast services (MBS) and unicast services along with the existing voice, internet, and other background traffic services over the wireless cellular networks, it is essential to efficiently handle the wireless bandwidth in order to ensure the admission of more number of calls in the system during the congested traffic condition, to maximize the overall service quality, and to maximize the revenue. Fixed bandwidth allocation for the MBS sessions has several disadvantages. Whenever the MBS sessions are provided with the fixed bandwidth and minimum qualities, then the qualities of the MBS videos are lowest even if the traffic condition is very low. Therefore, for this condition, bandwidth utilization is also low. Similarly when the MBS sessions are provided with the fixed bandwidth and maximum qualities, then the MBS videos are provided with the best qualities but due to the shortage of the wireless bandwidth for the non-MBS traffic calls (e.g., voice, unicast video, internet, and other background traffic), the overall forced call termination probability as well as the handover call dropping probability are increased. Also the revenue is decreased for the operators.

Scalable Video Coding (SVC) is an excellent solution to the problems raised by the diverse characteristics of high data rate video transmission through the wireless link. The SVC allows the elimination of some parts of the video bit stream in order to adapt it to the various needs or preferences of end users as well as to varying terminal capabilities or network conditions [9]. Therefore, scalable video technique [4, 9-11] allows the variable bit rate video broadcast/multicast/unicast over wireless networks. This technique utilizes multiple layering. Each of the layers improves spatial, temporal, or visual quality of the rendered video to the user [4]. Base layer or the highest priority layer guarantees the minimum quality of a video stream, whereas the addition of enhanced layers or low priority layers improves the video quality. The number of layers for a video session (program) and the bandwidth per layer can be manipulated dynamically. Thus, to broadcast/multicast/unicast videos through a wireless environment, layered transmission is an effective approach for supporting heterogeneous receivers with varying bandwidth requirements [11]. Hence, if the system bandwidth is not sufficient to allocate the demanded bandwidth for all of the broadcasting/multicasting/unicasting video sessions, it is possible to reduce the bandwidth allocation for each of the video sessions. The QoS adaptability [12-16] of some multimedia traffic types is also an important technique for wireless communication to increase the admission of more number of calls in the system. This technique can be applied to support more number of calls during the congested traffic condition for the MBS supported wireless cellular networks.

In this paper, we propose a bandwidth allocation scheme that efficiently allocates bandwidth among the MBS sessions and the non-MBS traffic calls. The proposed scheme reduces the bandwidth allocation for the MBS sessions during the congested traffic condition only to accommodate more non-MBS traffic calls in the system. Instead of allocating fixed bandwidths for the BMS sessions and non-MBS traffic, our scheme allocates variable bandwidths for them. However, the minimum quality of the videos is guaranteed by allocating minimum bandwidth for each of the video sessions. The SVC technique allows the reduced bandwidth allocation for the MBS sessions and the unicast videos. The proposed scheme also reduces the bandwidth allocation for the background traffic based on the QoS adaptability. Using the mathematical and numerical analyses, we show that the proposed scheme maximizes the bandwidth utilization and significantly reduces the overall forced call termination probability and the handover call dropping probability.

The rest of this paper is organized as follows. Section 2 introduces the system model of the proposed scheme. Bandwidth adaptation and bandwidth allocation procedures are described in Section 3. In Section 4, we provide the queuing model and formulas for the new call blocking probability and the handover call dropping probability. Performance evaluation results of the proposed scheme are presented and compared with other schemes in Section 5. Finally, Section 6 concludes our work.

## 2. System Model

The proposed scheme is based on the dynamic bandwidth allocation for the MBS sessions and non-MBS traffic calls. The allocated bandwidth for the MBS sessions is based on the system capacity and the condition of traffic congestion. For the lower traffic condition, the MBS sessions are provided with the maximum demanded bandwidth for each of the sessions. With the increase of the traffic arrival rates, the total allocated bandwidth for the MBS sessions is decreased. However, the minimum bandwidth is allocated for the MBS sessions to guarantee the minimum video qualities. The bandwidth allocation for the background traffic is also reduced with the increase of the traffic congestion. The bandwidth allocation for the unicast video calls is decreased only for highly congested traffic conditions. We consider non-MBS traffic (e.g., voice, unicast, and background) and MBS sessions. We also consider that the MBS sessions are always active and they are provided with at least minimum amount of bandwidth. We give priority for each of the traffic types. The handover calls of any types of calls are considered as highest priority calls. The next priority is given to voice and unicast video calls. The background traffic and the MBS sessions are given lowest priority. Fig. 1 shows the basic concept of bandwidth allocations for the MBS sessions and the non-MBS traffic calls. In the low traffic condition, all calls are provided with the maximum qualities. However, for the congested traffic condition, the bandwidth allocation for the MBS sessions is decreased. Based on the bandwidth allocation policy for different traffic calls, the system allocates bandwidth for the non-MBS traffic calls. Then, the remaining bandwidth is allocated for the MBS sessions. Suppose $C_{max,B}$ and $C_{min,B}$ are, respectively, the maximum allowable bandwidth and the minimum allocated bandwidth for the active MBS video sessions. $C_{max,nB}$ and $C_{min,nB}$ are, respectively, the maximum allowable bandwidth and the minimum allowable bandwidth for the non-MBS traffic calls. The bandwidth $C_{max,B}$ is provided to MBS sessions only if the allocated bandwidth for the non-MBS traffic calls is less than or equal to $C_{min,nB}$. Fig. 2 shows the variation of bandwidth allocations for the MBS sessions and the non-MBS traffic calls with the increase of demanded bandwidth by the non-MBS traffic calls. During the lower traffic condition, all the MBS sessions and non-MBS traffic calls are provided with the maximum demanded bandwidth for each of them. When the total demanded bandwidth exceeds the system capacity, the system reduces the bandwidth allocation both for the MBS sessions and non-MBS traffic calls. The reduction of bandwidth for the non-MBS traffic calls depends on the number of running QoS adaptive calls. The quality of each of the MBS sessions is guaranteed by allocated minimum amount of bandwidth for the MBS sessions.

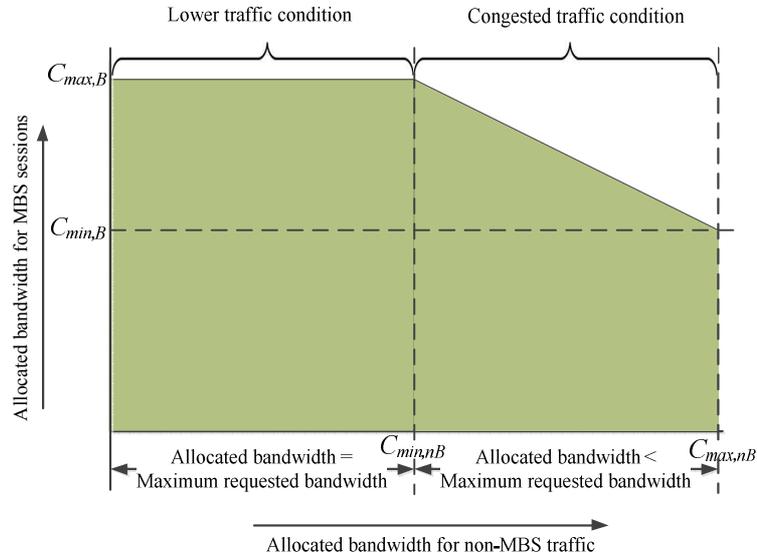

**Fig. 1** Basic concept of bandwidth allocations among the MBS sessions and the non-MBS traffic calls.

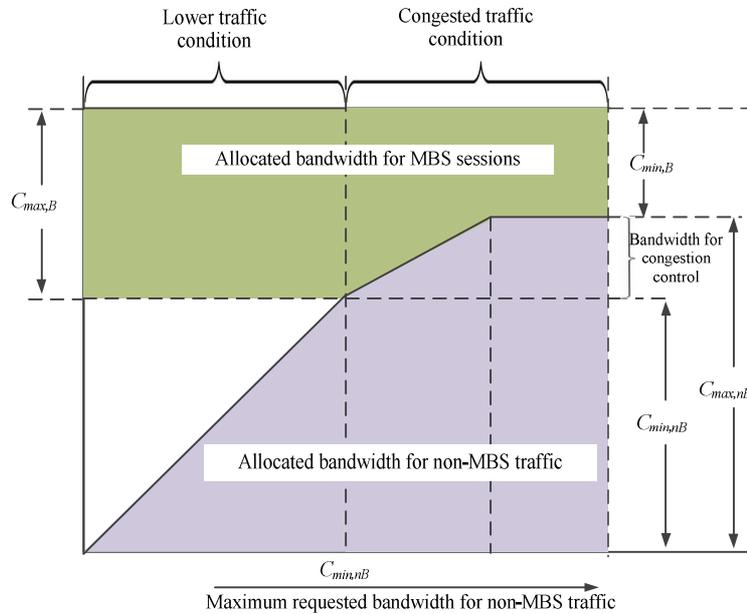

**Fig. 2** Variation of bandwidth allocations for the MBS sessions and the non-MBS traffic calls.

## 3. Proposed Bandwidth Allocation and Adaptation

Even though the bandwidth capacities of various wireless networks are growing very rapidly, fully deployed 4G wireless network will not be even enough to accommodate many best quality video services simultaneously with the other traffic. Efficient bandwidth allocation for video sessions in broadcasting/multicasting/unicasting and for other voice or data traffic is needed to make the best usage of the scare resources of wireless networks. An easy and straightforward approach is that all of the active broadcasting/multicasting video sessions and non-MBS traffic calls are provided by the requested bandwidth. However, such approach is not effective to serve huge traffic calls. Therefore, we propose an efficient bandwidth allocation scheme that makes the best utilization of the wireless bandwidth. The proposed scheme allows to reclaim some of the allocated bandwidth from already admitted QoS degradable traffic calls (e.g., background traffic and unicast video) and MBS sessions, as to accept more handover and new calls, when the system's resources are running low. Consequently, the scheme can accommodate more calls. Fig. 3 shows the procedure of bandwidth degradation for the proposed scheme. The proposed scheme gives highest priority for any kinds of handover calls. Suppose $C_{req,max}$ and $C_{req,min}$ are, respectively, the maximum and the minimum required bandwidths for a requested call. The system accommodates a handover call if it can manage $C_{req,min}$ amount of bandwidth only. However, for a new arrival call, it is equal to $C_{req,max}$. The qualities of the unicast video calls are degraded only to accept handover calls in the system. The overall resource allocation scheme is divided into four categories based on the traffic characteristics. The resource allocation and QoS adaptation for each of the traffic types are different.

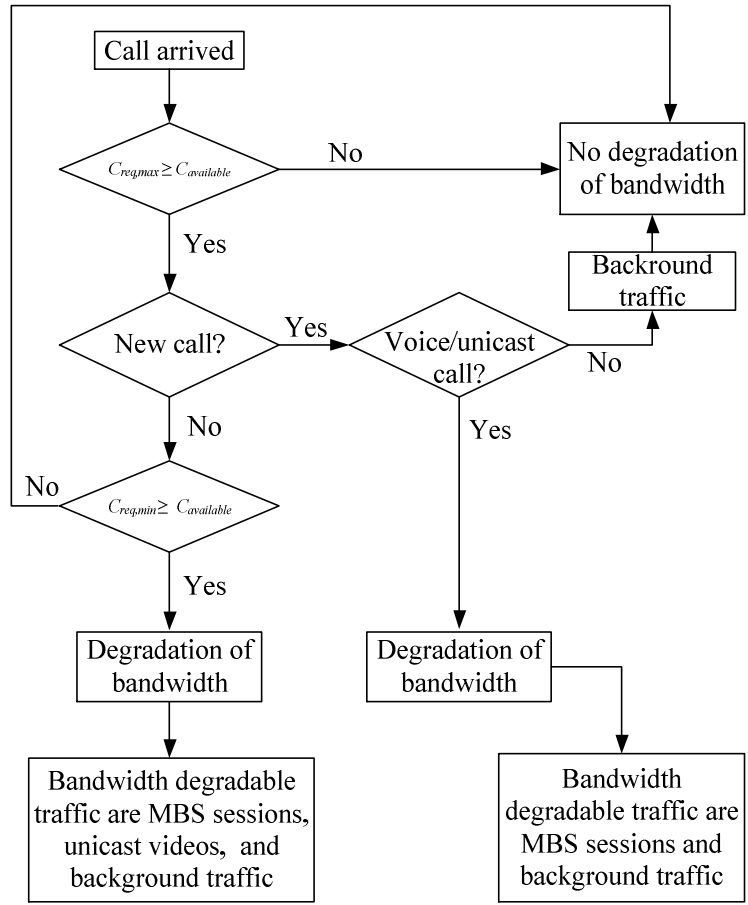

**Fig. 3** Procedure of bandwidth degradation of the system traffic.

### 3.1 Voice Traffic

The proposed scheme gives highest priority to the voice calls followed by the handover calls. The QoS of this traffic type is non-adaptive. Hence, the bandwidth allocation for the voice traffic is strict. However, the QoS levels of other classes of traffic are degraded to admit a voice call. Suppose $\beta_v$, $\beta_{min,v}$, and $\beta_{max,v}$ are, respectively, the currently allocated, minimum allocated, and maximum allocated bandwidths for a voice call. The bandwidth relation for the voice calls is found as:

$$\beta_v = \beta_{min,v} = \beta_{max,v} \tag{1}$$

### 3.2 Unicast Video (Scalable)

The unicast calls are also given same priority as the voice calls. The QoS of this traffic type is adaptive. The number of enhanced layers is reduced to accommodate more handover calls in the system. The QoS levels of other classes of traffic are also degraded to admit a unicast video call. The unicast video calls show the following characteristics for the proposed scheme.

- Same priority as the voice calls
- Multiple levels of bandwidth allocations including maximum bandwidth allocation for best quality (maximum number of enhanced layers) and minimum bandwidth allocation for guaranteed quality (minimum number of enhanced layers)
- The number of enhanced layers can be controlled to accommodate more handover calls in the system
- QoS level of MBS sessions and the background traffic calls are degraded to admit a unicast call

The bandwidth relations for the unicast video calls are expressed as follows:

$$\beta_{min,uni} = \beta_{0,uni} + \beta_{1,uni} + \cdots + \beta_{K_{min},uni} \qquad (2)$$

$$\beta_{max,uni} = \beta_{0,uni} + \beta_{1,uni} + \cdots + \beta_{K_{min},uni} + \cdots + \beta_{K_{max},uni} \qquad (3)$$

$$\beta_{uni} = \beta_{0,uni} + \beta_{1,uni} + \cdots + \beta_{K_{min},uni} + \cdots + \beta_{k,uni} \qquad (4)$$

where $\beta_{uni}$, $\beta_{min,uni}$, and $\beta_{max,uni}$ are, respectively, the currently allocated, minimum allocated, and maximum allocated bandwidths for a unicast video call. $\beta_{0,uni}$ is the allocated bandwidth for the base layer of a unicast video call. $K_{max}$ and $K_{min}$ are, respectively, the maximum and the minimum numbers of supported enhanced layers for each of the unicast video calls. $\beta_{k,uni}$ is the required bandwidth for the *k-th* layer of a unicast call.

### *3.3 Multicast/Broadcast Video (Scalable)*

The QoS of this traffic type is adaptive. The numbers of enhanced layers are reduced to accommodate more handover calls as well as new voice and unicast video calls. The MBS sessions show the following characteristics for the proposed scheme.

- Multiple number of videos are continuously broadcasted/multicasted
- Multiple levels of bandwidth allocations including maximum bandwidth allocation for best quality (maximum number of enhanced layers) and minimum bandwidth allocation for guaranteed quality (minimum number of enhanced layers)
- The number of enhanced layers is controllable to accommodate more unicast video calls (both the new originating and handover calls), voice calls (both the new originating and handover calls), and handover calls of background traffic
- Based on the priorities, different multicasting/broadcasting video sessions have different levels of bandwidth allocations

The bandwidth relations for the MBS video sessions are expressed as follows:

$$\beta_{min,m} = \beta_{0,m} + \beta_{1,m} + \cdots + \beta_{N_{min,m},m} \qquad (5)$$

$$\beta_{max,m} = \beta_{0,m} + \beta_{1,m} + \cdots + \beta_{N_{min,m},m} + \cdots + \beta_{N_{max,m},m} \qquad (6)$$

$$\beta_{B,m} = \beta_{0,m} + \beta_{1,m} + \cdots + \beta_{N_{min,m},m} + \cdots + \beta_{N_m,m} \qquad (7)$$

$$C_{min,B} = \beta_{min,1} + \beta_{min,2} + \cdots + \beta_{min,m} + \cdots + \beta_{min,M}$$
$$= \sum_{m=1}^{M} \beta_{0,m} + \sum_{m=1}^{M} \sum_{n=1}^{N_{min,m}} \beta_{n,m} \qquad (8)$$

$$C_{max,B} = \beta_{max,1} + \beta_{max,2} + \cdots + \beta_{max,m} + \cdots + \beta_{max,M}$$
$$= \sum_{m=1}^{M} \beta_{0,m} + \sum_{m=1}^{M} \sum_{n=1}^{N_{min,m}} \beta_{n,m} + \sum_{m=1}^{M} \sum_{n=N_{min,m}+1}^{N_{max,m}} \beta_{n,m} \quad (9)$$

where $\beta_{B,m}$, $\beta_{min,m}$, and $\beta_{max,m}$ are, respectively, the currently allocated, minimum allocated, and maximum allocated bandwidths for *m-th* MBS session. $\beta_{0,m}$ is the allocated bandwidth for the base layer of the *m-th* MBS session. $N_{max,m}$ and $N_{min,m}$ are, respectively, the maximum and the minimum numbers of supported enhanced layers for the *m-th* MBS session. $\beta_{n,m}$ is the required bandwidth for the *n-th* layer of the *m-th* MBS session. $M$ is the number of active MBS sessions.

If $C_B$ and $C_{nB}$ are, respectively, the allocated bandwidths for the MBS sessions and the non-MBS traffic calls, then the lower traffic condition is defined as, $C - C_{nB} \geq \sum_{m=1}^{M} \sum_{n=0}^{N_{max,m}} \beta_{n,m}$. For this condition, the allocated bandwidth for the non-MBS traffic calls is less than or equal to the $C_{min,nB}$. Therefore, all the MBS sessions are provided with the maximum allowable bandwidth. The allocated bandwidth for the MBS sessions during this traffic condition is calculated as:

$$C_B = C_{max,B} = \sum_{m=1}^{M} \sum_{n=0}^{N_{max,m}} \beta_{n,m} \quad (10)$$

$$\beta_{B,m} = \beta_{max,m} = \beta_{0,m} + \beta_{1,m} + \cdots + \beta_{N_{min,m},m} + \cdots + \beta_{N_{max,m},m} \quad (11)$$

Congested traffic condition is defined as, $C - C_{nB} < \sum_{m=1}^{M} \sum_{n=0}^{N_{max,m}} \beta_{n,m}$. For this condition, the allocated bandwidth for the non-MBS traffic call is greater than $C_{min,nB}$. Therefore, all the MBS sessions are not provided with the maximum allowable bandwidth. For the congested traffic condition, the allocated bandwidth for the MBS sessions is calculated using two separate proposed techniques. For each of these techniques, the total allocated bandwidths for the MBS sessions are equal. However, the allocated bandwidths for different active sessions are different.

*Technique 1 (Two level bandwidth degradation):* This technique is applicable when the video qualities of all the MBS sessions need to be equally degraded. The proposed scheme provides almost equal degradation of MBS video qualities. The maximum difference between the reduced numbers of enhanced layers for two MBS sessions is one. Therefore, if the reduced number of enhanced layers for the most popular video session is $P$, the reduced number of enhanced layers for the lowest popular video session is either $P$ or $(P+1)$. The bandwidth for each of the MBS sessions is calculated as:

$$C_B = \sum_{m=1}^{M_1} \sum_{n=0}^{N_{max,m}-P} \beta_{n,m} + \sum_{m=M_1+1}^{M} \sum_{n=0}^{N_{max,m}-P-1} \beta_{n,m} \quad (12)$$

$$\beta_{B,m} = \begin{cases} \beta_{0,m} + \beta_{1,m} + \cdots + \beta_{N_{min,m},m} + \cdots + \beta_{N_{max,m}-P,m}, & 1 \leq m \leq M_1 \\ \beta_{0,m} + \beta_{1,m} + \cdots + \beta_{N_{min,m},m} + \cdots + \beta_{N_{max,m}-P-1,m}, & M_1 < m \leq M \end{cases} \quad (13)$$

where $P$ is the minimum number of enhanced layers that must be removed from every active MBS sessions due to the congestion of traffic. $M_1$ is the minimum number of MBS sessions for which $P$ number of enhanced layers are removed, and for the remaining ($M-M_1$) number of MBS sessions ($P+1$) number of enhanced layers are removed.

$P$ is the only value that satisfies both (14) and (15), whereas $M_1$ is the only value that satisfies both (16) and (17).

$$\frac{C-C_{nB}}{\sum_{m=1}^{M}\sum_{n=0}^{N_{max,m}-P-1}\beta_{n,m}}\geq 1 \tag{14}$$

$$\frac{C-C_{nB}}{\sum_{m=1}^{M}\sum_{n=0}^{N_{max,m}-P}\beta_{n,m}}<1 \tag{15}$$

$$\frac{C-C_{nB}-\sum_{m=1}^{M}\sum_{n=0}^{N_{max,m}-P-1}\beta_{n,m}}{\sum_{m=1}^{M_1}\beta_{(N_{max,m}-P),m}}\geq 1 \tag{16}$$

$$\frac{C-C_{nB}-\sum_{m=1}^{M}\sum_{n=0}^{N_{max,m}-P-1}\beta_{n,m}}{\sum_{m=1}^{M_1+1}\beta_{(N_{max,m}-P),m}}<1 \tag{17}$$

*Technique 2 (Multi-level bandwidth degradation):* This technique is applicable when the MBS video qualities are provided according to the priority of each of the sessions. In this scheme, all of the enhanced layers for the lowest priority sessions are removed first. The bandwidth for the MBS sessions is calculated as:

$$C_B = \sum_{m=1}^{M_2}\sum_{n=0}^{N_{max,m}}\beta_{n,m} + \sum_{m=M_2+1}^{M}\sum_{n=0}^{N_{min,m}}\beta_{n,m} \tag{18}$$

$$\beta_{B,m} = \begin{cases} \beta_{0,m}+\beta_{1,m}+\cdots+\beta_{N_{min,m},m}+\cdots+\beta_{N_{max,m},m}, & 1\leq m\leq M_2 \\ \beta_{0,m}+\beta_{1,m}+\cdots+\beta_{N_{min,m},m}, & M_2<m\leq M \end{cases} \tag{19}$$

where $M_2$ is the minimum number of MBS sessions for which the system provides best quality, and for the remaining ($M$-$M_2$) number of MBS sessions, the system provides minimum quality.

$M_2$ is the only value that satisfies both (20) and (21).

$$\frac{C-C_{nB}}{\sum_{m=1}^{M}\sum_{n=0}^{N_{min,m}}\beta_{n,m}+\sum_{m=1}^{M_2}\sum_{N_{min,m}+1}^{N_{max,m}}\beta_{n,m}}\geq 1 \tag{20}$$

$$\frac{C-C_{nB}}{\sum_{m=1}^{M}\sum_{n=0}^{N_{min,m}}\beta_{n,m}+\sum_{m=1}^{M_2+1}\sum_{N_{min,m}+1}^{N_{max,m}}\beta_{n,m}}<1 \tag{21}$$

### 3.4 Background Traffic (e.g., file transfer)

The QoS of background traffic is adaptive. The QoS adaptability [12-16] of this traffic allows the reclaiming of system resources to support more number of higher priority calls without reducing the bandwidth utilization. Two levels of bandwidth adaptation for the background traffic calls are proposed. The first level (higher) is used to accommodate handover calls and the second one (lower) is used to accommodate more new calls in the system. The background traffic shows the following characteristics for the proposed scheme.
− Lowest priority of traffic
− QoS is adaptive
− QoS level is degradable to accommodate more unicast video calls ( both the new originating and handover calls), voice calls (both the new originating and handover calls), and handover calls of background traffic

The bandwidth relations for the background traffic are expressed as follows:

$$\beta_{min,back(i)} = \beta_{hand,back(i)} = (1-\xi_i)\beta_{r,back(i)} \qquad (22)$$

$$\beta_{max,back(i)} = \beta_{r,back(i)} \qquad (23)$$

$$\beta_{new,back(i)} = (1-\xi_i')\beta_{r,back(i)} \qquad (24)$$

where $\beta_{r,back(i)}$, $\beta_{min,back(i)}$, and $\beta_{max,back(i)}$ are, respectively, the requested, minimum allocated, and maximum allocated bandwidths for a background traffic call of *i-th* class. $\beta_{hand,back(i)}(\beta_{new,back(i)})$ is the minimum required bandwidth to accept a handover (new) background traffic call of *i-th* class or minimum allocated bandwidth for each of the background traffic calls of class *i* after accepting any handover (new) calls. $\xi_i$ and $\xi_i'$ are, respectively, the maximum levels of bandwidths that can be degraded for a background traffic call of *i-th* class to accept a handover call and a new call.

We may compare our proposed scheme with few other schemes where the allocated bandwidths for MBS sessions are fixed. Few possible bandwidth allocation schemes are summarized as follows:

*Scheme #1 (proposed scheme):*
 – Proposed dynamic bandwidth allocation for MBS sessions
 – Priority based proposed dynamic bandwidth allocation for non-MBS traffic calls

*Scheme #2:*
 – Fixed $C_{max,B}$ amount of bandwidth allocation for MBS sessions
 – Priority based proposed dynamic bandwidth allocation for non-MBS traffic calls

*Scheme #3:*
 – Fixed $C_{max,B}$ amount of bandwidth allocation for MBS sessions
 – QoS degradable but non-prioritized bandwidth allocation for non-MBS traffic calls

*Scheme #4:*
 – Fixed $C_{max,B}$ amount of bandwidth allocation for MBS sessions
 – Non-QoS degradable as well as non-prioritized bandwidth allocation for non-MBS traffic calls

*Scheme #5:*
 – Fixed $C_{min,B}$ amount of bandwidth allocation for MBS sessions
 – Priority based proposed dynamic bandwidth allocation for non-MBS traffic calls

*Scheme #6:*
 – Fixed $C_{min,B}$ amount of bandwidth allocation for MBS sessions
 – QoS degradable but non-prioritized bandwidth allocation for non-MBS traffic calls

*Scheme #7:*
 – Fixed $C_{min,B}$ amount of bandwidth allocation for MBS sessions
 – Non-QoS degradable as well as non-prioritized bandwidth allocation for non-MBS traffic calls

Scheme #2 - scheme #4 provide always best quality for all the MBS sessions even in the very congested traffic condition. For these schemes $C_{max,B}$ amount of bandwidth is reserved for the MBS sessions. This $C_{max,B}$ amount of bandwidth is able to provide best quality videos for all the MBS sessions. These schemes can provide better bandwidth utilization but cannot improve the system performance in terms of overall forced call termination probability due to the reduced bandwidth allocation for the non-MBS traffic calls. Among these three schemes, only the scheme #2 can moderately improve the handover call dropping probability performance due to the presence of the proposed priority scheme for the non-MBS traffic calls. The scheme #2 can also reduce the new call blocking probabilities for the voice and the unicast traffic calls because of the priority based admission control but these reduced new call blocking probabilities are not significant. For the scheme #3 and scheme #4, the handover call dropping probability is very high because of the non-prioritized call admission control. Scheme #4 provides lowest number of call admission in the system.

Scheme #5 - scheme #7 provide always lowest quality for all the MBS sessions even for the very low traffic condition. For these schemes, only $C_{min,B}$ amount of bandwidth is reserved for the MBS sessions. This $C_{min,B}$ amount of bandwidth is able to provide only the lowest quality videos for all the MBS sessions. These schemes cannot perform well in terms of bandwidth utilization for the lower traffic condition. In the lower traffic condition, even the bandwidth is empty but the MBS sessions are provided with the lowest qualities. Scheme #5 and scheme #6 can maximize the number of call admission because of the increased bandwidth for the non-MBS sessions and the presence of the QoS degradation policy for the non-MBS traffic calls. Among these three schemes, only the scheme #5 can significantly improve the handover call dropping probability performance due the increased bandwidth for the non-MBS sessions and the presence of the proposed priority based QoS degradation policy for the non-MBS traffic calls. The scheme #5 can also significantly reduce the new call blocking probabilities for the voice and the unicast traffic calls. For the scheme #6 and scheme #7, the handover call dropping probability is very high because of the non-prioritized call admission control.

Each of the schemes from scheme #2 – scheme #7 has some limitations. Therefore, we propose a new scheme (scheme #1) that optimizes all the limitations and provides efficient utilization of wireless bandwidth. Our scheme can provide maximum number of call admissions, significantly reduced handover call dropping probabilities, significantly reduced new call blocking probabilities for the voice and the unicast traffic calls, and maximized bandwidth utilization.

## 4. Queuing Model

Our proposed scheme can be analyzed using *M/M/K/K* queuing model. The resource allocation policy is varied with the increase of system states. Fig. 4 shows the state transition rate diagram for the proposed resource allocation scheme. We define $1/\mu$ as the average channel holding time (exponentially distributed). In this figure, $\lambda_n$ and $\lambda_h$ are, respectively, the total new call arrival rate and the handover call arrival rate. $\lambda_{n,x}$ is the new call arrival rate of *x* ( *v* or *uni* or *b*) type of traffic calls (*v* for voice, *uni* for unicast, and *b* for the background traffic calls). The *M* MBS sessions are continuously provided. Therefore, the minimum number of states in the system is *M*. The QoS degradation status for a traffic type is also obtainable form the state of the system.

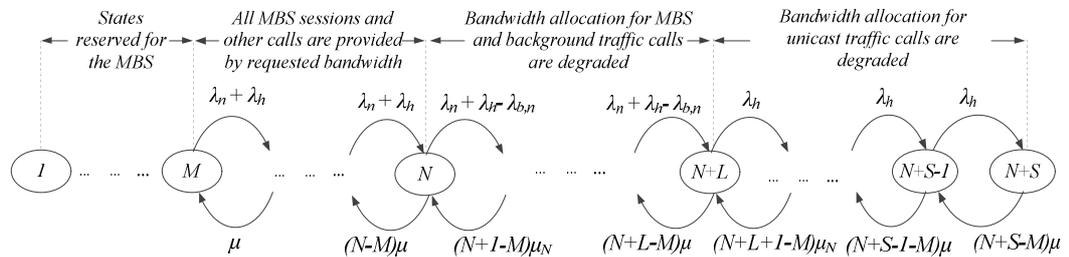

**Fig. 4.** State transition rate diagram for the proposed resource allocation scheme.

The system states can be divided into four parts to explain the resource allocation strategy.

***States 1 to M:***
- These states are reserved to serve *M* MBS sessions

***States 1 to N:***
- All calls are supported by the maximum demanded bandwidths
- All $N_{max,m}$ enhanced layers are supported for all the MBS sessions
- All $K_{max}$ enhanced layers are supported for the unicast services

***States N+1 to N+L:***
- Bandwidth allocation for the MBS sessions and background traffic calls are degraded to accommodate new calls of voice and unicast traffic, and all handover calls (voice, unicast, and background)
- All $K_{max}$ enhanced layers are supported for the unicast video services
- Number of enhanced layers for each of the MBS sessions are provided based on the priority of the video session or based on the bandwidth allocation technique for MBS sessions

***States N+L+1 to N+S:***
- MBS sessions and background traffic calls are already degraded their maximum possible limits
- Unicast traffic calls are degraded to accommodate handover calls (voice, unicast, and background)
- All $K_{max}$ enhanced layers are not supported for the unicast services
- Only minimum number of enhanced layers are supported for each of the MBS sessions

The probability that the system is in state *i* is given by $P_i$. From the Fig. 4, the state balance equations are expressed as (25). In the proposed scheme, a new voice call or a unicast video call is blocked if the system is in the state $(N+L)$ or larger and a new background traffic call is blocked if the system is in the state *N* or larger. However, a handover call is dropped only if the system is in the state $(N+S)$. Thus, the handover call dropping probability ($P_D$) is calculated using (26). The call blocking probability of an originating new voice call ($P_{B,v}$) or unicast video call ($P_{B,uni}$) can be computed using (27). Finally, the call blocking probability of an originating new background traffic call ($P_{B,back}$) can be computed using (28).

$$\begin{aligned}(i-M)\mu P_i &= \lambda_T P_{i-1}, & M \leq i \leq N \\ (i-M)\mu P_i &= \left(\lambda_{n,v} + \lambda_{n,uni} + \lambda_h\right)P_{i-1}, & N \leq i \leq N+L \\ (i-M)\mu P_i &= \lambda_h P_{i-1}, & N+L \leq i \leq N+S\end{aligned} \quad (25)$$

$$P_D = P_{N+S} = \frac{\lambda_T^{N-M}\left(\lambda_{n,v} + \lambda_{n,uni} + \lambda_h\right)^L \left(\lambda_h\right)^{S-L}}{\mu^{N+S-M}(N+S-M)!}P_M \quad (26)$$

$$P_{B,v} = P_{B,uni} = \sum_{i=N+L}^{N+S} P_i = \sum_{i=N+L}^{N+S} \frac{(\lambda_T)^{N-M}\left(\lambda_{n,v} + \lambda_{n,uni} + \lambda_h\right)^L \left(\lambda_h\right)^{i-N-L}}{\mu^{i-M}(i-M)!}P_M \quad (27)$$

$$P_{B,back} = \sum_{i=N}^{N+S} P_i = \sum_{i=N+L}^{N+S} P_i + \sum_{i=N}^{N+L-1} P(i) = P_{B,v} + \sum_{i=N}^{N+L-1}\frac{(\lambda_T)^{N-M}\left(\lambda_{n,v} + \lambda_{n,uni} + \lambda_h\right)^{i-N}}{\mu^{i-M}(i-M)!}P_M \quad (28)$$

$$\frac{1}{P_M} = \sum_{i=M}^{N}\frac{(\lambda_T)^{i-M}}{\mu^{i-M}(i-M)!} + \sum_{i=N+1}^{N+L}\frac{(\lambda_T)^{N-M}\left(\lambda_{n,v} + \lambda_{n,uni} + \lambda_h\right)^{i-N}}{\mu^{i-M}(i-M)!} + \sum_{i=N+L+1}^{N+S}\frac{(\lambda_T)^{N-M}\left(\lambda_{n,v} + \lambda_{n,uni} + \lambda_h\right)^L \lambda_h^{i-N-L}}{\mu^{i-M}(i-M)!}$$

$$(29)$$

# 5. Performance Evaluation

In this section, we present the results of the numerical analysis of the proposed scheme. We compare the performance of our proposed scheme with the performance of the "fixed bandwidth allocation for MBS sessions" schemes. Table 1 shows the assumptions of the parameter values used in analysis. The call arriving process and the cell dwell times are assumed to be Poisson. The average cell dwell time is assumed to be 540 sec (exponentially distributed) [17].

**Table 1** Summary of the parameter values used in analysis

| Parameter | Value |
|---|---|
| Bandwidth capacity ($C$) | 20 (Mbps) |
| Required bandwidth for each of the voice calls ($\beta_v$) | 64 (kbps) |
| Maximum allocated bandwidth for each of the unicast video calls ($\beta_{max,uni}$) | 0.5 (Mbps) |
| Maximum number of enhanced layers for each of the unicast video calls ($K_{max,uni}$) | 10 |
| Minimum number of enhanced layers for each of the unicast video calls ($K_{min,uni}$) | 0 |
| Allocated bandwidth for each of the enhanced layers of the unicast video calls ($\beta_{k,uni}$) | 20 (kbps) |
| Maximum allocated bandwidth for each of MBS sessions ($\beta_{max,m}$) | 1 (Mbps) |
| Minimum allocated bandwidth for each of MBS video sessions ($\beta_{min,m}$) | 0.5 (Mbps) |
| Maximum number of enhanced layers for each of MBS sessions ($N_{max,m}$) | 10 |
| Minimum number of enhanced layers for each of MBS sessions ($N_{min,m}$) | 0 |
| Allocated bandwidth for each of the enhanced layers of each of MBS sessions ($\beta_{n,m}$) | 50 (kbps) |
| Number of MBS sessions ($M$) | 12 |
| Maximum required/allocated bandwidth for each of background traffic calls ($\beta_{max,back}$) | 120 (kbps) |
| Minimum required/allocated bandwidth for each of background traffic calls ($\beta_{min,back}$) | 60 (kbps) |
| Maximum portion of bandwidth that can be degraded for a background traffic call of $i$-th class ($\xi_i$) | 0.5 |
| Maximum portion of bandwidth that can be degraded for a background traffic call of $i$-th class to accept a new originating call ($\xi'_i$) | 0.3 |
| Average call duration of non-MBS traffic calls considering all non-MBS traffic calls (exponentially distributed) | 120 (sec) |
| Ratio of call arrival rates (voice: unicast call: background traffic ) | 5:1:4 |

Fig. 5 shows the variation of bandwidth allocations for the MBS sessions and the non-MBS traffic calls with the increase of call arrival rate for our proposed scheme. It shows that the bandwidth allocation for the MBS sessions is decreased with the increase of demanded bandwidth by the non-MBS traffic calls. However, the minimum allocated bandwidth for the MBS sessions is 6 Mbps. Therefore, the maximum allocated bandwidth for the the non-MBS traffic calls is 14 Mbps. Fig. 6 shows the number of provided enhanced layers for a MBS session and a unicast video call with the increase of call arrival rate. It shows that the proposed system provides less quality degradation of unicast video calls compared to the MBS video sessions. The bandwidth allocation for the MBS sessions using "two level technique" causes maximum difference of one enhanced layer between the highest priority and the lowest priority sessions. The bandwidth allocation for the MBS sessions using "multi-level technique" causes difference of zero to ten enhanced layers between the highest priority and the lowest priority sessions.

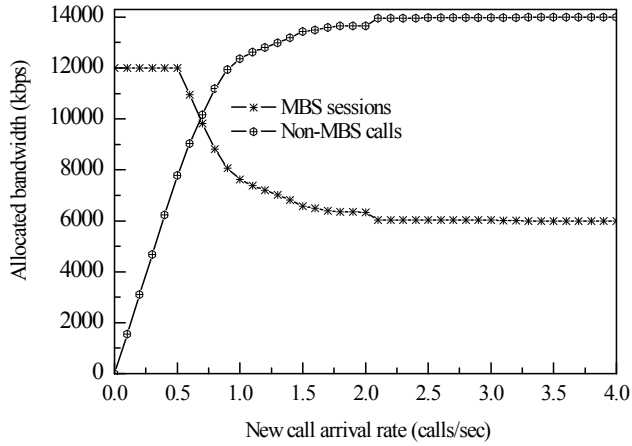

**Fig. 5** The variation of bandwidths allocations for the MBS sessions and the non-MBS traffic calls with the increase of call arrival rate.

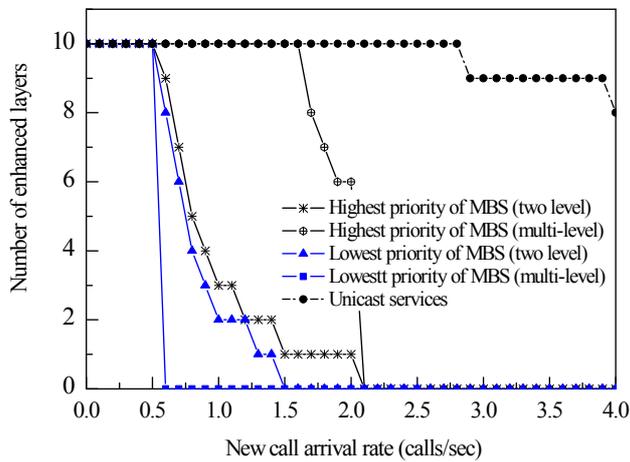

**Fig. 6** Number of provided enhanced layers for a MBS session and a unicast video call with the increase of call arrival rate.

Fig. 7 shows that the proposed scheme provides negligible handover call dropping probability even in very high traffic condition. The scheme #2 is also based on the QoS adaptation and priority of traffic classes. However, the reduced maximum bandwidth allocation for the non-MBS traffic calls causes higher handover call dropping probability. Even the QoS adaptation policy is applicable for the scheme #6 but it causes very high handover call dropping probability due the non-priority of traffic classes. Among the scheme #2 – scheme #7, only the scheme #5 can provide negligible handover call dropping probability but the MBS sessions are always provided with the minimum qualities for this scheme. The scheme #3 and scheme #6 cause very high handover call dropping probabilities and new call blocking probabilities for voice and unicast calls due the non-priority of traffic classes. The performance of scheme #4 is poorer than scheme #3's and performance of scheme #7 is poorer than scheme #6's in terms of handover call dropping probability for all traffic types and new call blocking probabilities for voice and unicast video calls because these two schemes do not support QoS adaptability and the priority of traffic classes. Fig. 8 shows that our proposed scheme provides comparatively lower new call blocking probabilities for the voice and unicast traffic calls. Our scheme provides only higher new call blocking probability for the background traffic calls but that is still lower than the scheme #2. The scheme #2 cannot significantly reduce the new call blocking probabilities for the voice and unicast traffic calls due to the reduced maximum bandwidth allocation for the non-MBS traffic calls.

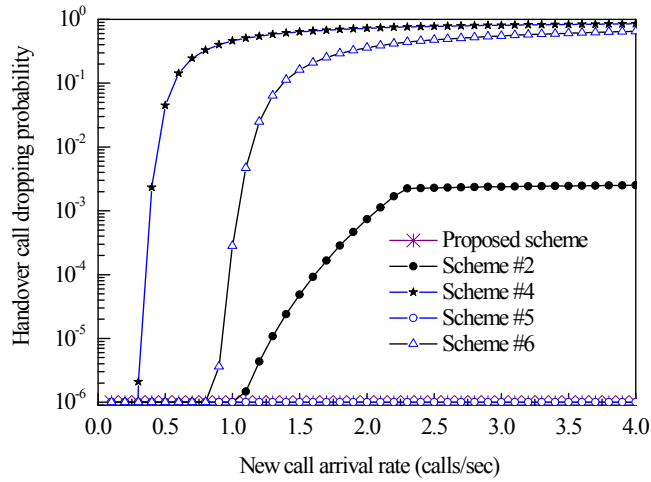

**Fig. 7** Comparison of handover call dropping probability.

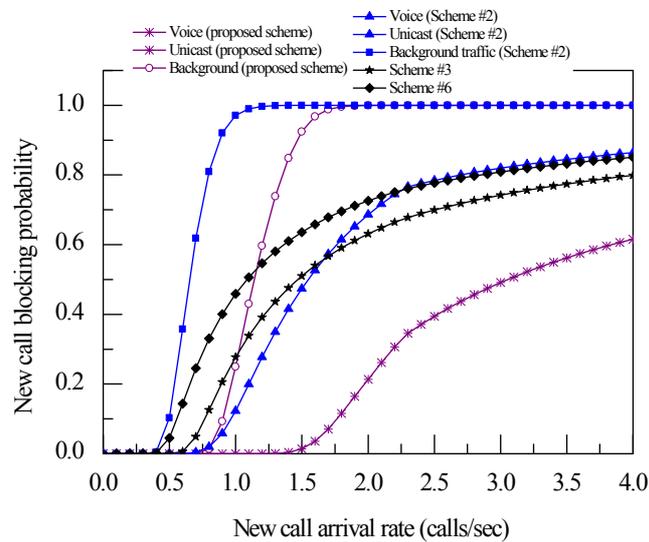

**Fig. 8** Comparison of new originating call blocking probability.

Fig. 9 shows the overall forced call termination probability performance comparison. Our proposed scheme provides best performance due to the dynamic nature of bandwidth allocation both for the MBS sessions and the non-MBS traffic calls. The scheme #2, scheme #3, and scheme #4 cannot improve the overall forced call termination performance due to the reduced maximum bandwidth allocation for the non-MBS traffic calls. Among the scheme #2 – scheme #7, only the scheme #5 and scheme #6 can maximize the number of call admission. However, the handover call dropping performance of the scheme #6 is poor. Scheme #7 performs poorer than scheme #6 and scheme #5 in terms of overall forced call termination probability because this scheme does not support QoS adaptability. During the lower traffic condition, the proposed scheme allocates higher bandwidth for the MBS sessions. The proposed scheme reduces the bandwidth allocation for the MBS sessions during the congested traffic condition only. Hence, our scheme effectively uses the system bandwidth. Fig. 10 shows the bandwidth utilization comparison. Even though the scheme #5 can maximize the number of call admission but bandwidth utilization is poor for this scheme especially for the lower traffic condition. Therefore, the proposed scheme maximizes bandwidth utilization also.

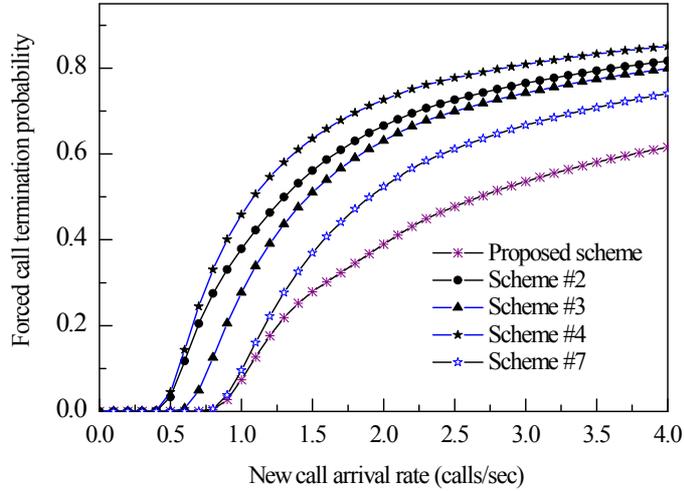

**Fig. 9** Comparison of overall forced call termination probability.

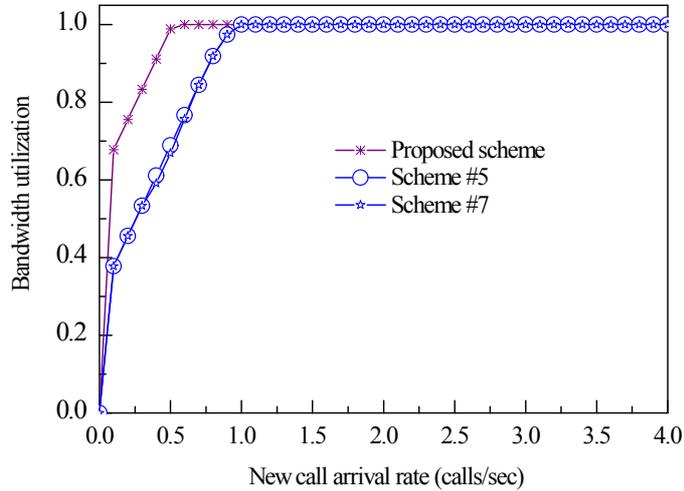

**Fig. 10** Comparison of bandwidth utilization.

The results in Figs. 7 - 10 show that the proposed scheme quite effective for the MBS services over the wireless networks. The only disadvantage of the proposed scheme is that the video quality of the MBS sessions and QoS level of non-MBS traffic calls are degraded during the congested traffic condition.

## 6. Conclusions and Future Research

In this paper, we proposed a QoS adaptive bandwidth allocation scheme for MBS supported wireless cellular networks. The idea behind the proposed scheme is that, during the shortage of bandwidth, the system releases some bandwidth from the MBS video sessions and other QoS adaptive calls, as to accommodate more calls in the system. More bandwidth is released to support handover calls over new calls. Also more bandwidth is released to support new voice and unicast video calls over new background traffic calls. Thus, the proposed scheme results in negligible handover call dropping probability for all traffic types and lower new call blocking probability for voice calls and unicast calls.

We have shown that the proposed scheme is very much effective in reducing the handover call dropping probability without sacrificing the bandwidth utilization. The proposed scheme reduces the allocated bandwidth for the MBS sessions and other QoS adaptive non-MBS traffic calls only in the congested traffic condition. While the proposed scheme blocks more new background traffic calls during the congested traffic condition. The proposed scheme also maximizes the bandwidth utilization and number of call admissions.

While employing the proposed scheme, the network operator has the opportunity to increase the revenue. Consequently, the proposed scheme is expected to be a considerable interest for MBS provision through wireless cellular networks. We studied research issues concerning the efficient resource allocation for the high data rate video services through wireless networks. The research results were studied using several numerical analyses. Experimental results for comparison to theory are saved for future research work. However, our proposed scheme provides a good basis for research as well as industry to implement high data rate video services through wireless networks.

**Acknowledgements** This work was supported by Electronics and Telecommunications Research Institute (ETRI), Korea. This work was also supported by the IT R&D program of MKE/KEIT [10035362, Development of Home Network Technology based on LED-ID].

# References

1. UDCAST WiMAX TV Home Page. [Online]. Available: http://www.udcast.com/solutions/OLD/udcast_solutions_tv_wimax.htm
2. Wang, J., Venkatachalam, M., & Fang, Y. (2007). System architecture and cross-layer optimization of video broadcast over WiMAX. *IEEE Journal on Selected Areas in Communications,* 25(4), 712-721.
3. Chowdhury, M. Z, Jang, Y. M., & Haas, Z. J. (2011). Cost-effective frequency planning for capacity enhancement of femtocellular networks. *Wireless Personal Communications*, 60(1), 83-104.
4. Sharangi, S., Krishnamurti, R., & Hefeeda, M. (2011). Energy-efficient multicasting of scalable video streams over WiMAX networks. *IEEE Transactions on Multimedia*, 13(1), 102-115.
5. Ding, J.-W., Chen, Z.-Y., Lee, W.-T., & Chen, W.-M. (2009). Adaptive bandwidth allocation for layered video streams over wireless broadcast channels. In *Proceedings of International Conference on Communications and Networking* (pp. 1-5).
6. Lee, J. H., Pack, S., Kwon, T., & Choi, Y. (2011). Reducing handover delay by location management in mobile WiMAX multicast and broadcast services. *IEEE Transactions on Vehicular Technology,* 60(2), 605-617.
7. Lin, Y.-B. (2001). A multicast mechanism for mobile networks. *IEEE Communications Letters,* 5(11), 450-452.
8. Lin, C.-T., & Ding, J.-W. (2006). CAR: A low latency video-on-demand broadcasting scheme for heterogeneous receivers. *IEEE Transaction on Broadcasting*, 52(3), 336-349.
9. Ding, J.-W., Deng, D.-J., Wu, T.-Y., & Chen, H.-H. (2010). Quality-aware bandwidth allocation for scalable on-demand streaming in wireless networks. *IEEE Journal on Selected Areas in Communications,* 28(3), 366-376.
10. Wang, Y., Chau, L.-P., & Yap, K.-H. (2010). Bit-rate allocation for broadcasting of scalable video over wireless networks. *IEEE Transactions on Broadcasting,* 56(3), 288-295.
11. Liu, J., Li, B., Hou, Y. T., & Chlamtac, I. (2004). On optimal layering and bandwidth allocation for multisession video broadcasting. *IEEE Transactions on Wireless Communications*, 3(2), 656-667.
12. Chowdhury, M. Z., Jang, Y. M., & Haas, Z. J. (2013). Call admission control based on adaptive bandwidth allocation for multi-class services in wireless networks. *IEEE/KICS Journal of Communications and Networks,* 15(1), 15-24.
13. Vergados, D. D. (2007). Simulation and modeling bandwidth control in wireless healthcare information systems. *SIMULATION*, 83(4), 347–364.

14. Zhuang,W., Bensaou, B., & Chua, K. C. (2000). Adaptive quality of service handoff priority scheme for mobile multimedia networks. *IEEE Transactions on Vehicular Technology*, 49(2), 494-505.
15. Cruz-Perez, F. A., & Ortigoza-Guerrero, L. (2004). Flexible resource allocation strategies for class-based QoS provisioning in mobile networks. *IEEE Transactions on Vehicular Technology*, 53(3), 805-819.
16. Habib, I., Sherif, M., Naghshineh, M., & Kermani, P. (2003). An adaptive quality of service channel borrowing algorithm for cellular networks. *International Journal of Communication Systems,* 16(8), 759–777.
17. Schwartz, M. (2005). *Mobile Wireless Communications,* Cambridge University Press, Cambridge.

# Biographies

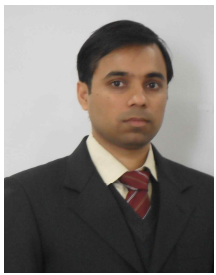

**Mostafa Zaman Chowdhury** received his B.Sc. degree in electrical and electronic engineering from Khulna University of Engineering and Technology (KUET), Bangladesh, in 2002. He received his M.Sc. and Ph.D. degrees both in electronics engineering from Kookmin University, Korea, in 2008 and 2012, respectively. In 2003, he joined the Electrical and Electronic Engineering Department at KUET as a faculty member. In 2008, he received the Excellent Student Award from Kookmin University. One of his papers received the Best Paper Award at the International Conference on Information Science and Technology in April 2012 in Shanghai, China. He served as a reviewer for several international journals (including IEEE Communications Magazine, IEEE Transaction on Vehicular Technology, IEEE Communications Letters, IEEE Journal on Selected Areas in Communications, Wireless Personal Communications (Springer), Wireless Networks (Springer), Mobile Networks and Applications (Springer), and Recent Patents on Computer Science) and IEEE conferences. He has been involved in several Korean government projects. His research interests include convergence networks, QoS provisioning, mobility management, femtocell networks, and VLC networks.

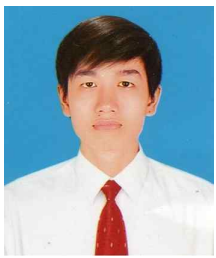

**Tuan Nguyen** received his B.Sc. degree in electronics and telecommunications from Hanoi University of Science and Technology (HUST), Vietnam, in 2011. Currently he is working towards his M.Sc. degree in the department of Electronics Engineering at Kookmin University, Korea. He has been involved in several Korean government projects. His research interests include convergence networks, VLC networks, LED-ID networks, localization, and QoS provisioning

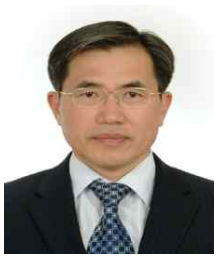

**Young-Il Kim** received the BS and MS degrees both in electronics engineering from Kyung Hee University, Korea, in 1985 and 1988, respectively. He received the doctoral degree in electronics engineering from Kyung Hee University, Korea, in 1996. He worked for Samsung Electronics between 1985 and 1986. Since 1988, he is with ETRI, Daejeon, Korea where currently he is working as a team leader of Smart screen convergence research department.

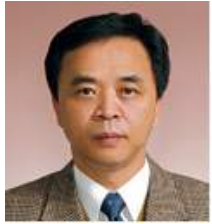

**Won Ryu** received the BS degree in computer science and statistics from Pusan National University, Busan, Korea, in 1983, and the MS degree in computer science and statistics from Seoul National University, Seoul, Korea, in 1988. He received his PhD degree in information engineering from Sungkyunkwan University, Kyonggi, South Korea, in 2000. Since 1989, he has been a managing director with the Smart screen convergence research department, ETRI, Daejeon, Korea. Currently, his research interests are IPTV, Smart TV, IMT-advanced, and convergence services and networks.

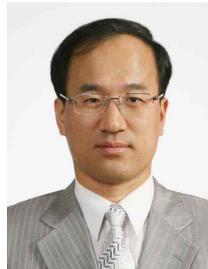

**Yeong Min Jang** received the B.E. and M.E. degrees both in electronics engineering from Kyungpook National University, Korea, in 1985 and 1987, respectively. He received the doctoral degree in Computer Science from the University of Massachusetts, USA, in 1999. He worked for ETRI between 1987 and 2000. Since September 2002, he is with the School of Electrical Engineering, Kookmin University, Seoul, Korea. He has organized several conferences such as ICUFN2009, ICUFN2010, ICUFN2011, ICUFN2012, and ICUFN2013. He is currently a member of the IEEE and a life member of KICS (Korean Institute of Communications and Information Sciences). He had been the director of the Ubiquitous IT Convergence Research Center at Kookmin University since 2005 and the director of LED Convergence Research Center at Kookmin University since 2010. He has served as the executive director of KICS since 2006. He had been the organizing chair of Multi Screen Service Forum of Korea since 2011. He had been the Chair of IEEE 802.15 LED Interest Group (IG-LED). He received the Young Science Award from the Korean Government (2003 to 2005). He had served as the founding chair of the KICS Technical Committee on Communication Networks in 2007 and 2008. His research interests include 5G mobile communications, radio resource management, small cell networks, multi-screen display networks, LED communications, ITS, and WPANs.